\documentclass[preprint,
superscriptaddress,
amsmath,amssymb,aps,showkeys,showpacs,
twoside,final,secnumarabic,
nofootinbib]{revtex4-2}

\usepackage[paperwidth=205mm,paperheight=290mm,top=17mm,bottom=25mm,
inner=17mm,outer=17mm,
twoside]{geometry}

\usepackage{cmap} 
\usepackage[T1,T2A]{fontenc}
\usepackage[utf8]{inputenc}
\usepackage[russian,english]{babel}
\usepackage{color}
\usepackage{graphicx}
\usepackage{dcolumn}
\usepackage{bm} 
\usepackage[unicode=true,colorlinks=true,linkcolor=magenta, urlcolor=blue, citecolor = blue,breaklinks]{hyperref}
\usepackage{multirow}
\usepackage{url}
\usepackage{breakurl}
\DeclareGraphicsExtensions{.eps}

\newcount\issue
\newcount\Vol
\newcount\numb
\usepackage{fancyhdr} 
\def\Vol{\textbf{80}}
\def\numb{x}
\usepackage{booktabs,bm}

\begin{document}

\title{ CONFERENCE SECTION \\[20pt]\boldmath
Analysis of $B\to KM_X$ and $B\to K^* M_X$ decays\\in scalar- and vector-mediator dark-matter scenarios}

\def\addressa{D.~V.~Skobeltsyn Institute of Nuclear Physics,\\M.~V.~Lomonosov Moscow State University, 119991, Moscow, Russia}
\def\addressb{Institute of High Energy Physics, Austrian Academy of Sciences,\\Nikolsdorfergasse 18, A-1050 Vienna, Austria}
\def\addressc{D.~V.~Skobeltsyn Institute of Nuclear Physics,\\M.~V.~Lomonosov Moscow State University, 119991, Moscow, Russia;\\Joint Institute for Nuclear Research, 141980 Dubna, Russia;\\ Faculty of Physics, University of Vienna, Boltzmanngasse 5, A-1090 Vienna, Austria}

\author{\firstname{Alexander}~\surname{Berezhnoy}}
\email[E-mail: ]{A.V.Berezhnoy@gmail.com}
\affiliation{\addressa}
\author{\firstname{Wolfgang}~\surname{Lucha}}
\email[E-mail: ]{Wolfgang.Lucha@oeaw.ac.at}
\affiliation{\addressb}
\author{\firstname{Dmitri}~\surname{Melikhov}}
\email[E-mail: ]{dmitri_melikhov@gmx.de}
\affiliation{\addressc}

\received{xx.xx.2025}
\revised{xx.xx.2025}
\accepted{xx.xx.2025}

\begin{abstract}
The surprising excess of missing-energy events far beyond all standard-model expectations in the~weak decays of the charged ground-state $B^+$ meson into some charged strange meson, rather recently observed by the Belle-II experiment, may (easily) be explained by the decay of the $B$ meson into the strange~meson and a pair of dark-matter fermion and antifermion, mediated by an (intermediate) scalar or vector~boson. Thorough inspections of both the total and the differential widths of these decays provide, among others, a simple means for the (straightforward) discrimination of such mediator boson's scalar or vector nature.
\end{abstract}

\pacs{13.20.He, 95.30.Cq, 95.35.+d}\par
\keywords{Dark Matter, mediator bosons, charged B-meson decays, missing-energy decays, Belle-II excess events\\[5pt]}

\maketitle
\thispagestyle{fancy}


\section{Motivation: Belle-II's Surprising Missing-Energy Excess Finding}

Comparatively recently, the Belle-II collaboration reported \cite{Belle-II:23} its observation of a (somewhat) astounding excess of decay events of the charged $B^+$ meson into a charged $K^+$ meson and missing energy $M_X$ (the latter also denoted a neutrino-antineutrino pair $\bar\nu\nu$), $B^+\to K^+\bar\nu\nu$. The outcome of such measurements may be quantified by contrasting the branching ratio of the missing-energy decay, ${\cal B}(B^+\to K^+\bar\nu\nu)$, with the branching ratio ${\cal B}(B^+\to K^+\bar\nu\nu)_{\rm SM}$ obtained from the standard model (SM) of particle physics for the $B^+$ decay into $K^+$ plus a neutrino-antineutrino pair,~$\bar\nu\nu$~\cite{Belle-II:23}:
\begin{equation}
{\cal B}(B^+\to K^+\bar\nu\nu)=(2.3\pm0.7)\cdot10^{-5}=(5.4\pm1.5)\,{\cal B}(B^+\to K^+\bar\nu\nu)_{\rm SM}\ .\label{ME}
\end{equation}
Interestingly, this figure \cite{Belle-II:23,Epi24,EPJST,WIFAI23,HQL23,QCD24,ICHEP24} is in \emph{slight} tension with an
upper limit published earlier by~Belle~\cite{Belle17},
$${\cal B}(B^+\to K^+\bar\nu\nu)<1.9\cdot10^{-5}\ .$$

Subsequent to the Belle II announcement of the surprising experimental finding (\ref{ME}) we~started (like many others) to investigate the evident possibility that the noticed excess of missing-energy events is a consequence of additional contributions by dark matter (DM) to the $B$-meson~decay. Instead of scrutinizing specific features of any DM environment, we focus \cite{BM,BLM1,BLM2} to the a little bit more general analysis of classes of models wherein the ``communication'' between~the~SM~particle sector and the DM particle sector is established by means of the exchange of mediator~bosons $R$ of either scalar, $R=\phi$, or vector, $R=V$, nature. In this context, of particular importance~proves to be a concurrent investigation of missing-energy $B^+$-meson decays to a pseudoscalar $K^+$ meson ($B^+\to K^+M_X$), on the one hand, or to a vector $K^{*+}$ meson ($B^+\to K^{*+}M_X$), on~the~other~hand.

For each $B^+$ decay of the kind mentioned above, the missing energy $M_X$ encompasses the total energy of all those decay products involved that ``succeed'' to escape their experimental~detection; accordingly, the missing energy $M_X$ recorded in this decay is fixed by the difference $q\equiv p_B-p_{K^{(*)}}$ of the $B$-meson momentum $p_B$ and the experimentally accessible $K^{(*)}$-meson's momentum, $p_{K^{(*)}}$:
$$M_X=\sqrt{q^2}\equiv\sqrt{(p_B-p_{K^{(*)}})^2}\ .$$

\begin{figure}[b]\includegraphics[width=7.271cm]{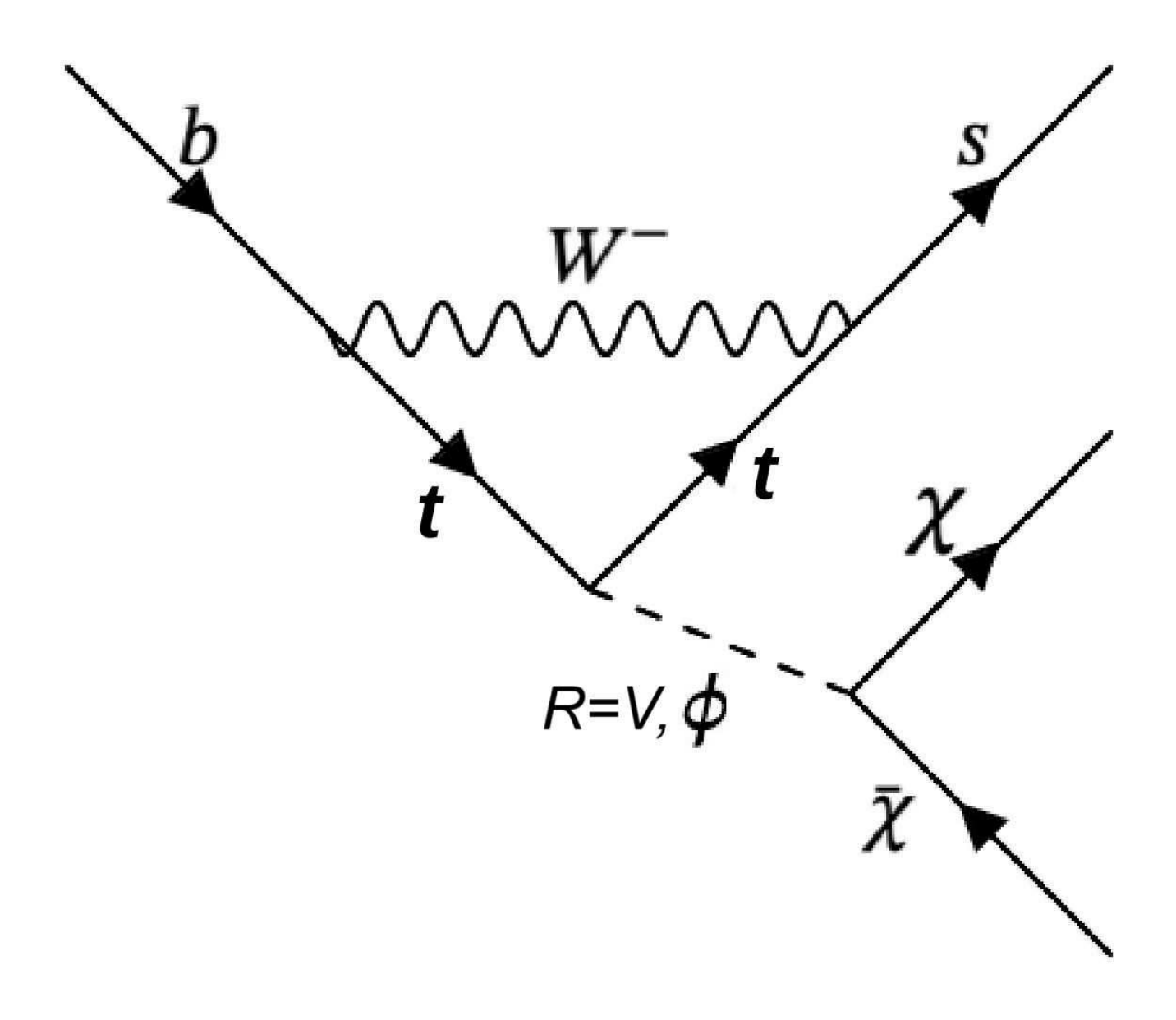}
\caption{\label{fig:penguin}Penguin diagram \cite{BLM2} depicting exchanges of a scalar ($\phi$) or vector ($V$) mediator $R$ [dashed line] between an SM top quark $t$ [solid lines] and a DM fermion $\chi$ [solid lines], contributing by an exchange of a charged SM gauge boson $W^-$ to the flavour-changing neutral-current process $b\xrightarrow{R}s\bar\chi\chi$ and eventually to the $B$ meson decay $B\xrightarrow{R}K^{(*)}\bar\chi\chi$ to a $K^{(*)}$ meson and a pair $\bar\chi\chi$ of DM fermion--antifermion~under~study.}
\end{figure}

For definiteness, let's assume that the mediator $R$ predominantly couples to an SM top~quark $t$ as well as to some DM fermions, $\chi$. This opens up the possibility of $B$-meson decays $B\xrightarrow{R}K^{(*)}\bar\chi\chi$ to $K^{(*)}$ mesons and DM fermion pairs $\bar\chi\chi$, enabled by the penguin-type process sketched in Fig.~\ref{fig:penguin}.
With the further missing-energy $B$-meson decay channels at our disposal, the missing~energy $M_X$ may manifest not only in form of the momenta $p_\nu$ of SM neutrinos $\nu$ and $p_{\bar\nu}$ of SM antineutrinos~$\bar\nu$,
$$q\equiv p_B-p_{K^{(*)}}=p_\nu+p_{\bar\nu}\qquad\Longrightarrow\qquad M_X^2=q^2\equiv(p_B-p_{K^{(*)}})^2=(p_\nu+p_{\bar\nu})^2\ ,$$
but, in addition, also in form of the momenta $p_\chi$ of a DM fermion $\chi$ and $p_{\bar\chi}$ of a DM antifermion~$\bar\chi$,
$$q\equiv p_B-p_{K^{(*)}}=p_\chi+p_{\bar\chi}\qquad\Longrightarrow\qquad M_X^2=q^2\equiv(p_B-p_{K^{(*)}})^2=(p_\chi+p_{\bar\chi})^2\ .$$
The widths governing these two types of decay channels $B\to K^{(*)}\bar\nu\nu$ and $B\xrightarrow{R}K^{(*)}\bar\chi\chi$ add~up~to
$$\Gamma(B\to K^{(*)}M_X)=\Gamma(B\to K^{(*)}\bar\nu\nu)_{\rm SM}+\Gamma(B\xrightarrow{R}K^{(*)}\bar\chi\chi)\ ,\qquad R=\phi,V\ .$$

\section{Dark-Matter Scenarios Involving Scalar or Vector Mediators}\label{sec:DM}

Beyond doubt, the evidently first move in any game of the kind intended should be to~identify all relevant quantities residing close to or at the interface of the SM sector and the DM sector and, subsequently, to determine or, at least, narrow down the numerical values of all these~parameters. To the latter quantities belong all \emph{involved} coupling strengths, the mass $m_\chi$ of the DM fermion~$\chi$, as well as the mass $M_R$ and the (potentially nonvanishing) decay width of the mediator $R=\phi,V$.

For the mediator mass $M_R$ larger than twice the dark-fermion mass $m_\chi$, that is, if $M_R>2\,m_\chi$, this mediator boson $R$ acquires a finite decay width to a dark fermion $\chi$ and a dark antifermion $\bar\chi$, which --- since entering in the full propagator of the mediator $R$ --- may be easily inferred from this mediator boson's two-point correlation function. The thus derived (by assumption predominant) decay width involves a $q^2$-independent factor $\Gamma^0_R$, governed by the coupling parameter $g_{R\chi\chi}$ of this mediator to a dark fermion--antifermion pair and multiplied by a factor exhibiting $q^2$~dependence.

We apply our ideas \cite{BM,BLM1,BLM2} to specific models for ``top-philic'' scalar ($\phi$) or vector ($V$) mediators.
\begin{itemize}
\item 
In a popular scalar-mediator model \cite{BPR11,SH13}, the interaction Lagrangian involves, on the~one hand, the SM coupling of $\phi$ to the top quark $t$ (usually formulated in terms of the $t$ mass $m_t$, the SM Higgs-boson vacuum expectation value $v~(\cong246~\mbox{GeV})$, and some parameter $y$) and, on the other hand, the DM coupling of $\phi$ to the dark fermion $\chi$ with coupling~strength~$g_{\phi\chi\chi}$:
\begin{equation}
\mathcal{L_\phi}=-\frac{y\,m_t}{v}\,\bar tt\,\phi-g_{\phi\chi\chi}\,\bar\chi\chi\,\phi\ .\label{LSMS}
\end{equation}
The loop in Fig.~\ref{fig:penguin} may be reduced to an \emph{effective} vertex with quark--mediator coupling~$g_{bs\phi}$:
\begin{equation}
\mathcal{L}_{b\rightarrow s\phi}=g_{bs\phi}\,\bar s_L b_R\,\phi\ .\label{LbsS}
\end{equation}
In the $\bar\chi\chi$ decay width of the mediator $\phi$, its $q^2$-independent factor $\Gamma^0_\phi$ is obtained~as~\cite{GS,MNNP}
$$\Gamma_\phi^0=\frac{g^2_{\phi\chi\chi}}{8\pi}\,M_\phi\left(1-\frac{4m_\chi^2}{M_\phi^2}\right)^{\frac{3}{2}}\ .$$
\item
Vector mediators $V$ may couple to vector and axialvector quark and dark-fermion currents. One specific model \cite{L08,CMRS,HLL} adopts as interaction Lagrangian (with couplings $g_{Vtt}$ and $g_{V\chi\chi}$)
\begin{equation}
\mathcal{L}_V=g_{Vtt}\,\bar t\gamma_\mu(1+\gamma_5)t\,V^\mu+g_{V\chi\chi}\,\bar\chi\gamma_\mu\chi\,V^\mu\ .\label{LSMV}
\end{equation}
The loop in Fig.~\ref{fig:penguin} ``collapses'' \cite{IL} to the \emph{effective} vertex with quark--mediator coupling~$g_{bsV}$
\begin{equation}
\mathcal{L}_{b\rightarrow sV}=g_{bsV}\,\bar s\gamma_\mu(1-\gamma_5)b\,V^\mu\ .\label{LbsV}
\end{equation}
For the mediator $V$, the $q^2$-independent term $\Gamma^0_V$ of the $\bar\chi\chi$ decay width is found~to~read~\cite{MNNP}
$$\Gamma_V^0=\frac{g^2_{V\chi\chi}}{12\pi}\,M_V\,\sqrt{1-\frac{4m_\chi^2}{M_V^2}}\left(1+\frac{2m_\chi^2}{M_V^2}\right).$$
\end{itemize}

\section{Experimental Discrimination of Scalar from Vector Mediator}

The actual nature of the possible mediator boson(s) $R$ --- scalar or vector (or maybe even~both) --- is of understandably great interest. In this context, our thorough analyses \cite{BM,BLM1,BLM2} stimulated by the $B^+\to K^+\bar\nu\nu$ branching-ratio observations (\ref{ME}) of Belle II \cite{Belle-II:23} underscore the importance of the simultaneous consideration of the two $R$-enabled missing-energy decays of a (charged)~$B^+$~meson
\begin{equation}
B\xrightarrow{R}K^{(*)}\bar\chi\chi\ ,\qquad K^{(*)}=K^+,K^{*+}\ ,\qquad R=\phi,V\ ,\label{BXX}
\end{equation}
to either the pseudoscalar $K^+$ meson or the vector $K^{*+}$ meson, and identify two (experimentally accessible) quantities of utmost relevance for the clarifications aimed at. The two expressions~are
\begin{enumerate}
\item the ratio of the differential decay widths, $d\Gamma/dq^2$, for the two missing-energy $B^+$ decays~(\ref{BXX}),
\begin{equation}
\widehat{\cal R}^{(R)}_{K^*/K}(q^2)\equiv
\dfrac{\;\dfrac{d\Gamma(B\xrightarrow{R}K^*\bar\chi\chi)}{dq^2}\;}{\dfrac{d\Gamma(B\xrightarrow{R}K\bar\chi\chi)}{dq^2}}\ ,\qquad R=\phi,V\ ;\label{Rdd}
\end{equation}
\item the ratio of the related integrated decay widths, $\Gamma$, for the two missing-energy $B$ decays~(\ref{BXX}),
\begin{equation}
{\cal R}^{(R)}_{K^*/K}\equiv\frac{\Gamma(B\xrightarrow{R}K^*\bar\chi\chi)}{\Gamma(B\xrightarrow{R}K\bar\chi\chi)}\ ,\qquad R=\phi,V\ .\label{Ri}
\end{equation}
\end{enumerate}

Straightforwardly, any hadronic $B\xrightarrow{R}K^{(*)}$ amplitude involved in these $B$-meson decay~widths may be expressed in terms of a couple of (dimensionless) form factors \cite{WSB}. For these form~factors, some convenient parametrizations \cite{MS} are available that encompass the nonperturbative~impacts of QCD by embedding outcomes of either lattice-QCD \cite{FL&M} or light-cone sum-rule \cite{BSZ,KMPW}~studies.

The (admittedly somewhat lengthy) explicit expressions of the requested differential $\bar\chi\chi$~decay widths $d\Gamma(B\xrightarrow{R}K^{(*)}\bar\chi\chi)/dq^2$, for $K^{(*)}=K^+,K^{*+}$, entering in the ratio (\ref{Rdd}) as well as (implicitly) in the \emph {integrated} decay widths forming the ratio (\ref{Ri}) can be found, for the case of a scalar mediator ($R=\phi$), in Refs.~\cite{BLM1,BLM2} and, for the case of a vector mediator ($R=V$), in Ref.~\cite{BLM2},~respectively.

For both scalar-mediator and vector-mediator case, the DM-sector related parameters --- more precisely, the couplings $g_{R\chi\chi}$ and $g_{bsR}$ in the interaction Lagrangians (\ref{LSMS}) or (\ref{LSMV}) and in the effective Lagrangians (\ref{LbsS}) or (\ref{LbsV}), respectively, the dark fermion's mass $m_\chi$, as well as the mass $M_R$ (and the $\bar\chi\chi$ decay width) of the mediator $R$ --- enter in the differential decay widths $d\Gamma(B\xrightarrow{R}K^{(*)}\bar\chi\chi)/dq^2$ by means of a factor that is common to both $d\Gamma(B\xrightarrow{R}K\bar\chi\chi)/dq^2$ and $d\Gamma(B\xrightarrow{R}K^*\bar\chi\chi)/dq^2$. That common factor necessarily \emph{cancels exactly} in the differential decay-width ratio (\ref{Rdd}). Consequently, this ratio (\ref{Rdd}) is absolutely insensitive to the (numerical values of any) DM parameters mentioned: Such differential decay-width ratio (\ref{Rdd}) does not exhibit any dependence on some DM parameters.

\newpage

\begin{center}\begin{figure}[hbt]\begin{tabular}{cc}\includegraphics[width=8.3544cm]{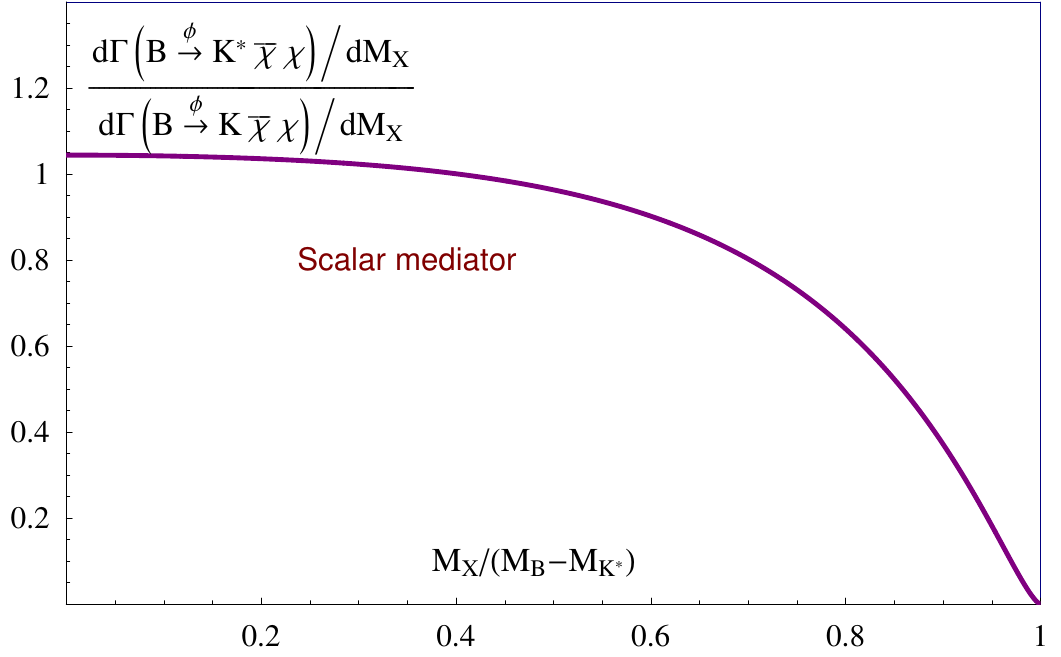}\hspace{0cm}&\hspace{.0cm} \includegraphics[width=8.3544cm]{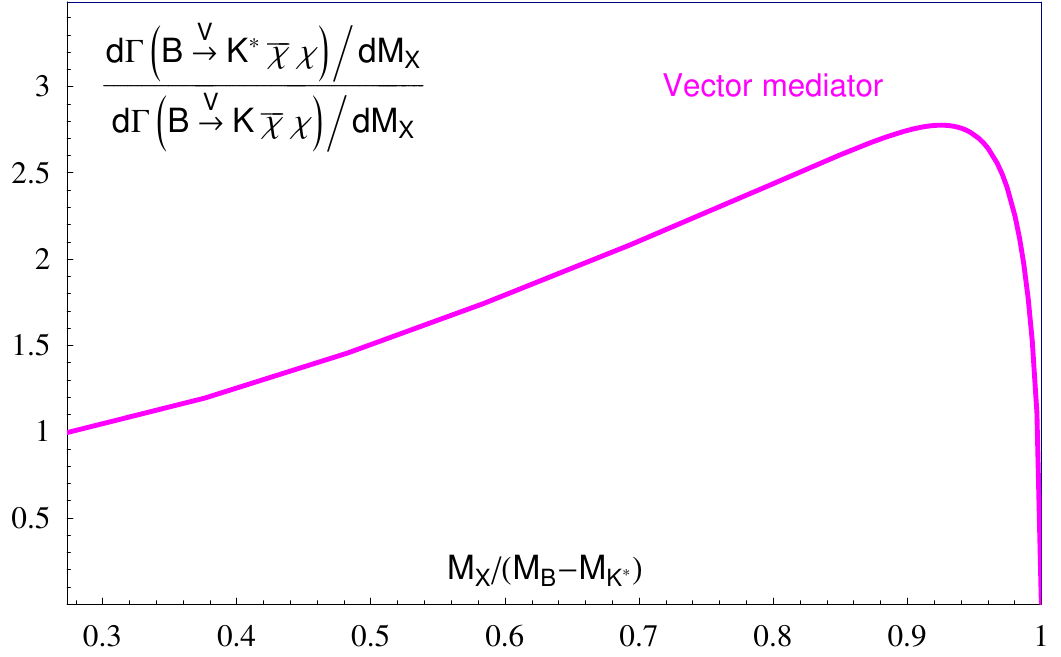}\\(a)&(b)\end{tabular}\caption{\label{Fig:Rd}Ratios (\ref{Rdd}) of the differential widths of $B^\pm$-meson decays into either $K^*$ or $K$ and a pair of dark fermions, as a function of the \emph{normalized} missing mass, $M_X$, for both scalar (a) and vector~(b)~mediators.}\end{figure}\end{center}

In striking contrast to the above insights on its total lack of dependence on the DM \emph{parameters} (for a really trivial reason), both differential decay-width ratios (\ref{Rdd}) turn out to be highly sensitive to the --- either scalar or vector --- \emph{nature} of any mediator boson $R$ effectuating the missing-energy decays $B\xrightarrow{R}K^{(*)}\bar\chi\chi$ in question, in other words, to the very spin of any participating mediator~$R$, as (unmistakably) illustrated in Fig.~\ref{Fig:Rd}: For scalar mediators $R=\phi$, the ratio (\ref{Rdd}) drops slowly~but monotonously from $\widehat{\cal R}^{(\phi)}_{K^*/K}(0)\gtrapprox1$ to its inevitable vanishing at $M_X/(M_B-M_{K^*})=1$. For~vector mediators $R=V$, this ratio (\ref{Rdd}) first grows from $\widehat{\cal R}^{(V)}_{K^*/K}(0)=1$ to its maximum (at comparatively large $M_X/(M_B-m_{K^*})\approx0.93$) and then drops rather sharply to its zero at $M_X/(M_B-M_{K^*})=1.$ This provides us with a tool for unambiguous identifications of adequate categories of DM~models by narrowing down the number of viable candidates (thus satisfying one of the aims of our study).

\begin{center}\begin{figure}[hbt]\begin{tabular}{cc}\includegraphics[width=8.3544cm]{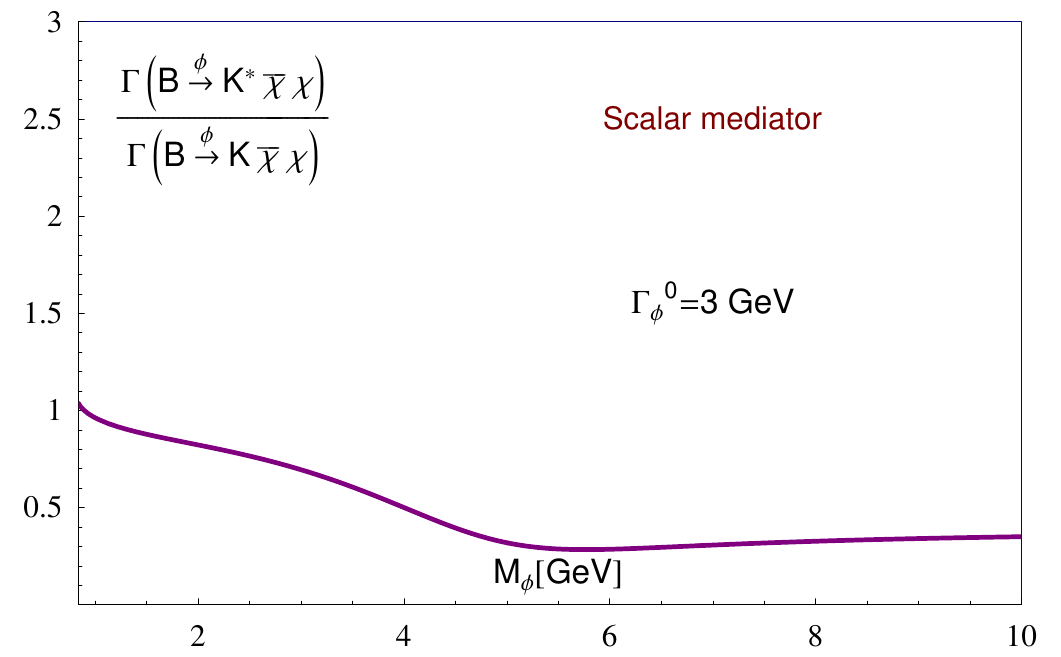}\hspace{0cm}&\hspace{0cm} \includegraphics[width=8.3544cm]{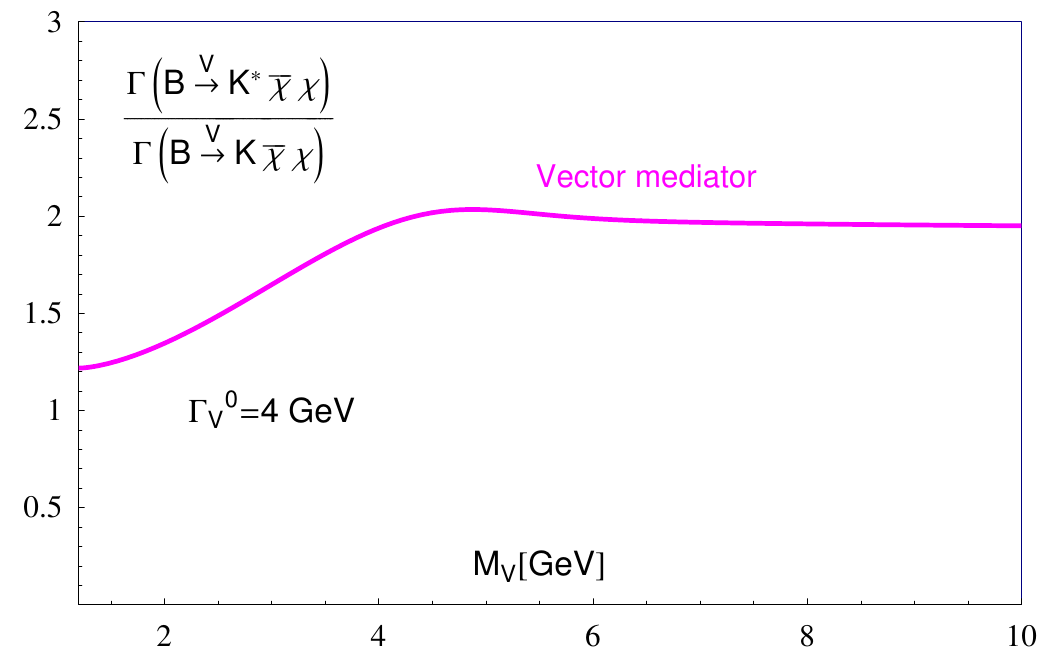}\\(a)&(b)\end{tabular}\caption{\label{Fig:Ri}Ratios (\ref{Ri}) of the integrated widths of $B^\pm$-meson decays into either $K^*$ or $K$ and a pair of dark fermions as a function of the mediator mass $M_R$, $R=\phi,V$, for both scalar (a) and vector~(b)~mediators.}\end{figure}\end{center}

The information on the DM sectors extractable from the ratio (\ref{Ri}) of integrated decay widths~$\Gamma$ constitutes, in a certain sense, a useful complement to the information provided in this respect by the differential decay-width ratio (\ref{Rdd}). In particular, the two integrated decay-width ratios~${\cal R}^{(R)}_{K^*/K}$ show a pronounced dependence on the mass $M_R$ --- and to a lesser extent on the decay width~$\Gamma^0_R$~--- of the mediator $R=\phi,V$, convincingly illustrated by their behaviour as functions of $M_R$ (Fig.~\ref{Fig:Ri}):
\begin{itemize}
\item In the case of \emph{scalar mediators}, for arbitrary mediator masses the ratio (\ref{Ri}) remains~below~1,
$${\cal R}^{(\phi)}_{K^*/K}(M_R)\lessapprox1\qquad\forall\ M_\phi\ .$$
\item In the case of \emph{vector mediators}, for arbitrary mediator masses the ratio (\ref{Ri}) is greater~than~1,
$${\cal R}^{(V)}_{K^*/K}(M_R)>1\qquad\forall\ M_V\ .$$
\end{itemize}

Still further enlightening conclusions can be gained upon performing a subtle expansion~of~the ratio of the missing-energy $B\to K^*M_X$ decay width and the (SM-fixed) $B\to K^*\bar\nu\nu$ decay~width,
\begin{equation}
\widetilde{\cal R}^{(R)}_{K^*}\equiv\frac{\Gamma(B\to K^*M_X)}{\Gamma(B\to K^*\bar\nu\nu)_{\rm SM}}\ ,\label{K*:ME/SM}
\end{equation}
in the form of a product of various, either theoretically or experimentally well-established~factors:
$$\widetilde{\cal R}^{(R)}_{K^*}=1+{\cal R}^{(R)}_{K^*/K}\,\frac{\Gamma(B\to K\bar\chi\chi)}{\Gamma(B\to K\bar\nu\nu)_{\rm SM}}\,\frac{\Gamma(B\to K\bar\nu\nu)_{\rm SM}}{\Gamma(B\to K^*\bar\nu\nu)_{\rm SM}}\ .$$
Inserting in the latter factorization of the ratio (\ref{K*:ME/SM}), for its third factor, the SM predictions~\cite{TW1,TW2,TW3}
\begin{eqnarray*}
{\cal B}(B^+\to K^+\bar\nu\nu)&=&(4.44\pm0.30)\cdot10^{-6}\ ,\\
{\cal B}(B^+\to K^{*+}\bar\nu\nu)&=&(9.8\pm1.4)\cdot10^{-6}\ ,
\end{eqnarray*}
and, for its second factor, our $\bar\chi\chi$ ``interpretation'' emerging from the Belle-II \cite{Belle-II:23} measurement~(\ref{ME})
$$\Gamma(B^+\to K^+\bar\chi\chi)=(4.4\pm1.5)\,\Gamma(B^+\to K^+\bar\nu\nu)_{\rm SM}\ ,$$
enables us to gain an idea of the relationship between $\bar\chi\chi$-vs.-SM ratio (\ref{K*:ME/SM}) and $K^*$-vs.-$K$ ratio~(\ref{Ri}):
\begin{equation}
\widetilde{\cal R}^{(R)}_{K^*}=1+(2\pm 0.6)\,{\cal R}^{(R)}_{K^*/K}\ .\label{K*:ME/SM(R)e}
\end{equation}
Furthermore, if merging the Belle upper bound on the branching ratio of the decay $B\to K^*\bar\nu\nu$~\cite{Belle17}
$${\cal B}(B\to K^*\bar\nu\nu)<2.7\cdot 10^{-5}$$
with the predicted branching ratio of the decay $B^+\to K^{*+}\bar\nu\nu$ and taking into account the~latter's estimated uncertainty, we are in the position to establish a rigorous upper bound on the ratio~(\ref{K*:ME/SM}):
\begin{equation}
\widetilde{\cal R}^{(R)}_{K^*}<3.2\ .\label{K*:ME/SM-UB}    
\end{equation}

\newpage

\begin{center}\begin{figure}[ht]\begin{tabular}{cc}\includegraphics[width=8.3544cm]{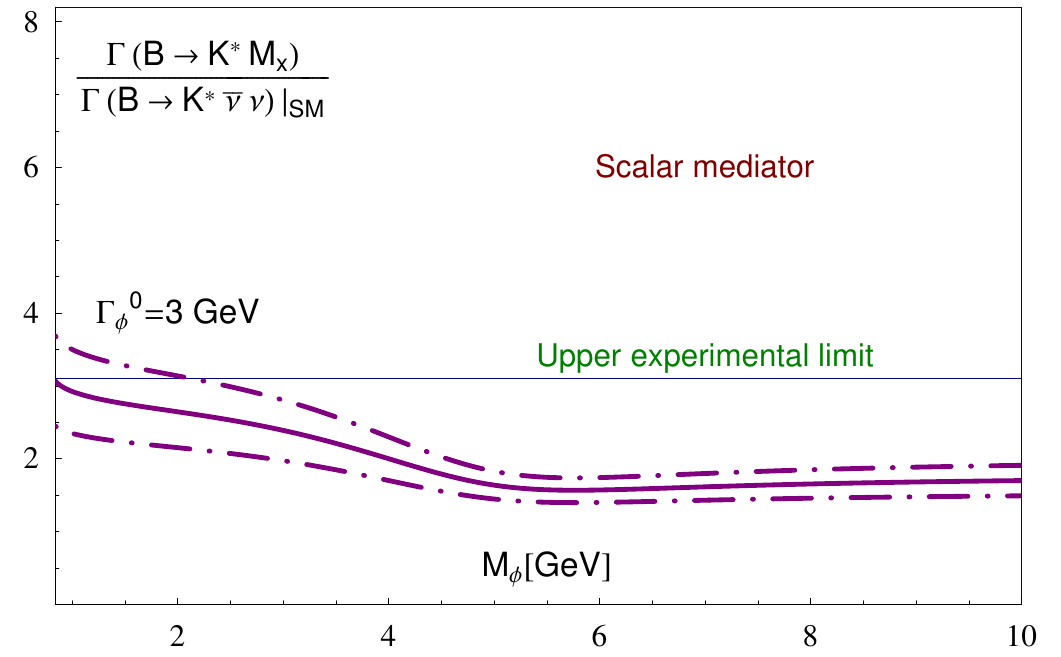}\hspace{0cm}&\hspace{0cm} \includegraphics[width=8.3544cm]{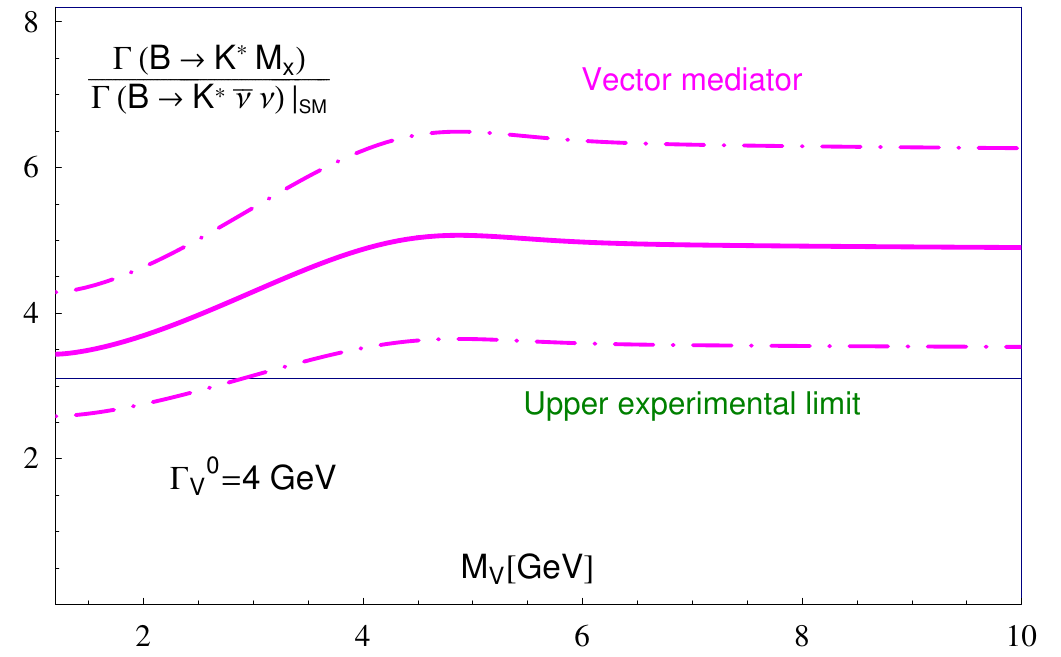}\\(a)&(b)\end{tabular}\caption{\label{Fig:EUB}Ratios (\ref{K*:ME/SM}) of integrated widths as a function of the mediator mass $M_R$ ($R=\phi,V$): dependence (\ref{K*:ME/SM(R)e}) on the ratio (\ref{Ri}) of integrated DM widths [solid line] with errors [dot-dashed line] confronted with the experimentally governed upper bound (\ref{K*:ME/SM-UB}) [horizontal line], for both scalar (a) and vector (b)~mediators.}\end{figure}\end{center}

Requiring mutual compatibility of the behaviour (\ref{K*:ME/SM(R)e}) of the ratio (\ref{K*:ME/SM}) with the (experimentally driven) upper bound (\ref{K*:ME/SM-UB}) might help impose even more constraints on the nature of the mediator, i.e., $R=\phi$ or $R=V$, and on some DM parameters. While such a compatibility request~turns out to have marginal impact for a scalar mediator $\phi$, it acts almost prohibitively for a vector~mediator $V$, by confining (as betrayed by Fig.~\ref{Fig:EUB}) the mass $M_V$ of the vector mediator to a rather low~range:
\begin{equation}
M_V\lesssim3\mbox{ GeV}\label{MV-UB}\ .
\end{equation}

\section{\label{mystery}Belle-II Missing-Energy Mystery: Dark Matter Interpretations}

The ultimate task of our study is to demonstrate that the $q^2$ distribution of the missing-energy events measured by Belle II \cite{Belle-II:23} can be reproduced by the two DM scenarios exemplified in~Sect.~\ref{sec:DM}. Experimental boundary conditions \cite{Belle-II:23} necessitate to use, instead of the variable $q^2$ adopted~so~far, some ``reconstruction'' variable, $q^2_{\rm rec}$, in terms of \emph{adequate} $B$ and $K$ energies $E_B$ and $E_K$ defined~by
$$q^2_{\rm rec}=E_B^2+M_K^2-2\,E_B\,E_K\ .$$

\begin{center}\begin{figure}[ht]\begin{tabular}{cc}
\includegraphics[width=8.354cm]{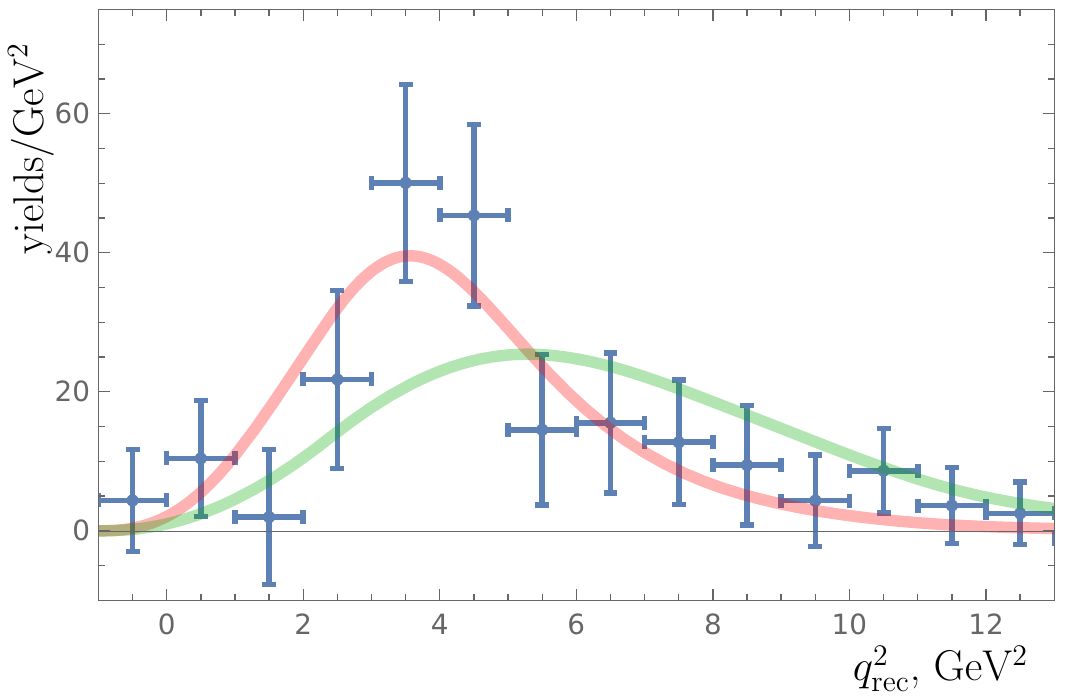}\hspace{0cm}&\hspace{0cm}
\includegraphics[width=8.354cm]{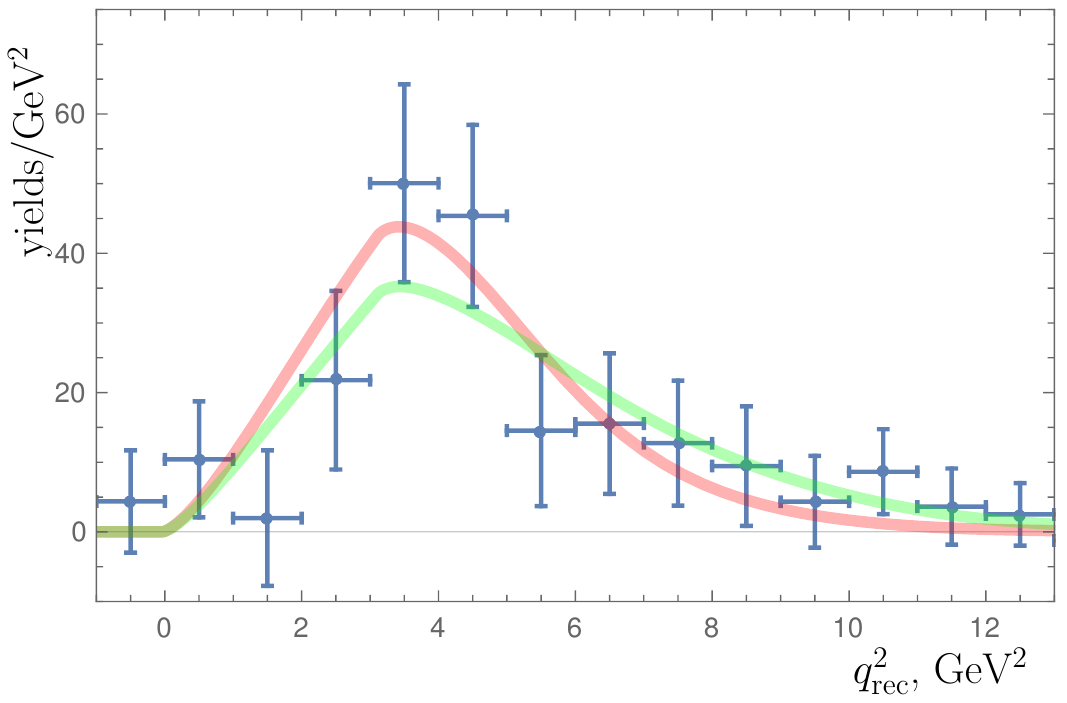}\\(a)&(b)\end{tabular}\caption{\label{Fig:fit}Fit of our generic DM models for scalar (a) and vector (b) mediators to Belle-II data \cite{Belle-II:23,FGOT}~on the decay $B\to KM_X$; red solid curve: global $\chi^2$-minimum yielding the DM parameter~values~in Table~\ref{ParVal+}; green solid curve: large mediator-mass examples at $M_\phi=20\mbox{ GeV}$, $\Gamma^0_\phi=20\mbox{ GeV}$ and $m_\chi=0.42\mbox{ GeV}$ for scalar mediators, resp.\ $M_V=20\mbox{ GeV}$ (excluded!), $\Gamma_V^0=4\mbox{ GeV}$ and $m_\chi=0.6\mbox{ GeV}$ for vector~mediators.} \end{figure}\end{center}

The actual feasibility of achieving, within both scalar- and vector-mediator settings, definitely acceptable fits to the $q^2_{\rm rec}$ distribution of missing-energy events presented by Belle II \cite{Belle-II:23} --- and this even if respecting the upper bound (\ref{MV-UB}) on the vector mediator's mass $M_V$ by fixing $M_V$ (a priori) to the value $M_V=3\mbox{ GeV}$ --- is underscored by overlaying, in Fig.~\ref{Fig:fit}, the Belle-II~outcomes~\cite{Belle-II:23} with the respective fit results. All thereby derived numerical values of the involved DM parameters~are collected in form of Table~\ref{ParVal+}. In order to ascertain that this amount of agreement need not~happen necessarily, Fig.~\ref{Fig:fit} also shows the implications of opting for less realistic values of DM parameters.

\begin{table}[ht]\caption{\label{ParVal+}Numerical values of (basic) mediator and DM-fermion mass and coupling parameters governing the scalar-mediator and vector-mediator DM models sketched in Sect.~\ref{sec:DM}, determined by our minimum-$\chi^2$ fits (as represented by the red solid curves in Fig.~\ref{Fig:fit}) to the Belle-II result \cite{Belle-II:23,FGOT} on the decay $B\to KM_X$.}
\begin{center}\begin{tabular}{lcc}\toprule
\textbf{Crucial Model Parameters}&$\hspace{1.104ex}$\textbf{Scalar Mediator $R=\phi$} \cite{BLM1}$\hspace{1.104ex}$&\textbf{Vector Mediator $R=V$} \cite{BLM2}\\\midrule
mediator mass $M_R$ [GeV]&$2.4\pm0.4$&$3$\\
mediator width $\Gamma^0_R$ [GeV]&$2.9^{+1.1}_{-0.9}$&$4.0^{+2.0}_{-1.5}$\\
dark-fermion mass $m_\chi$ [GeV]&$0.42^{+0.2}_{-0.4}$&$0.6^{+0.10}_{-0.18}$\\
mediator--DM-fermion coupling $g_{R\chi\chi}$&$\approx6$&$\approx7$\\
SM-quark--mediator coupling $g_{bsR}$&$\approx5\cdot10^{-8}$&$\approx2\cdot10^{-8}$
\\[1ex]\bottomrule\end{tabular}\end{center}\end{table}

\begin{center}\begin{figure}[ht]\begin{tabular}{cc}
\includegraphics[width=8.354cm]{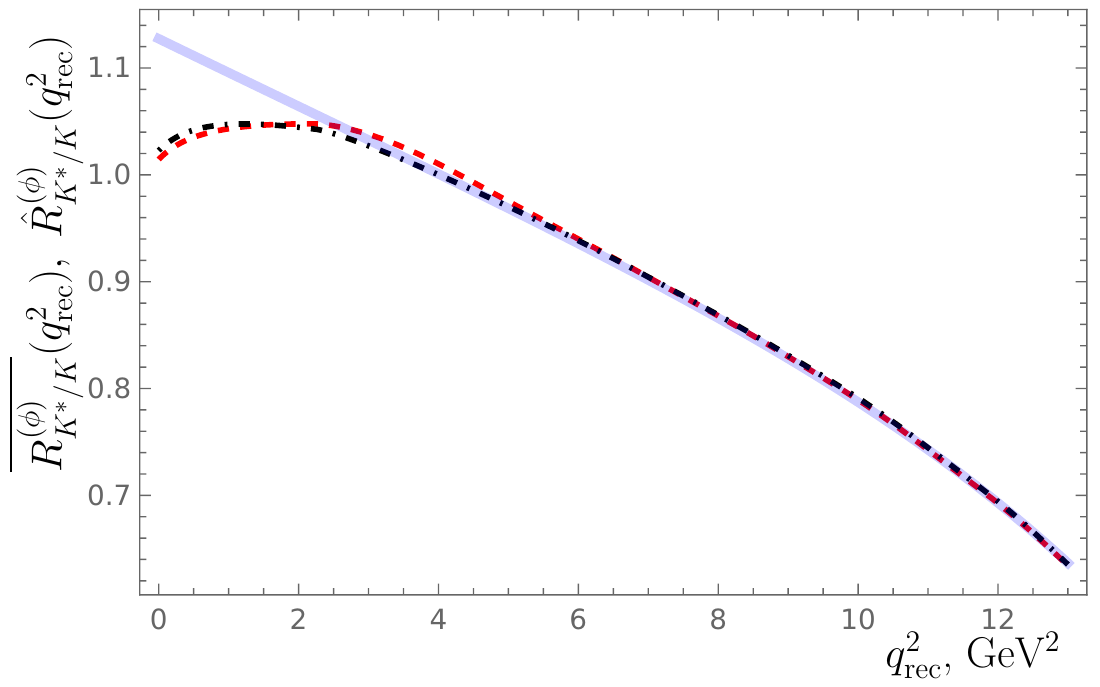}\hspace{0cm}&\hspace{0cm} \includegraphics[width=8.354cm]{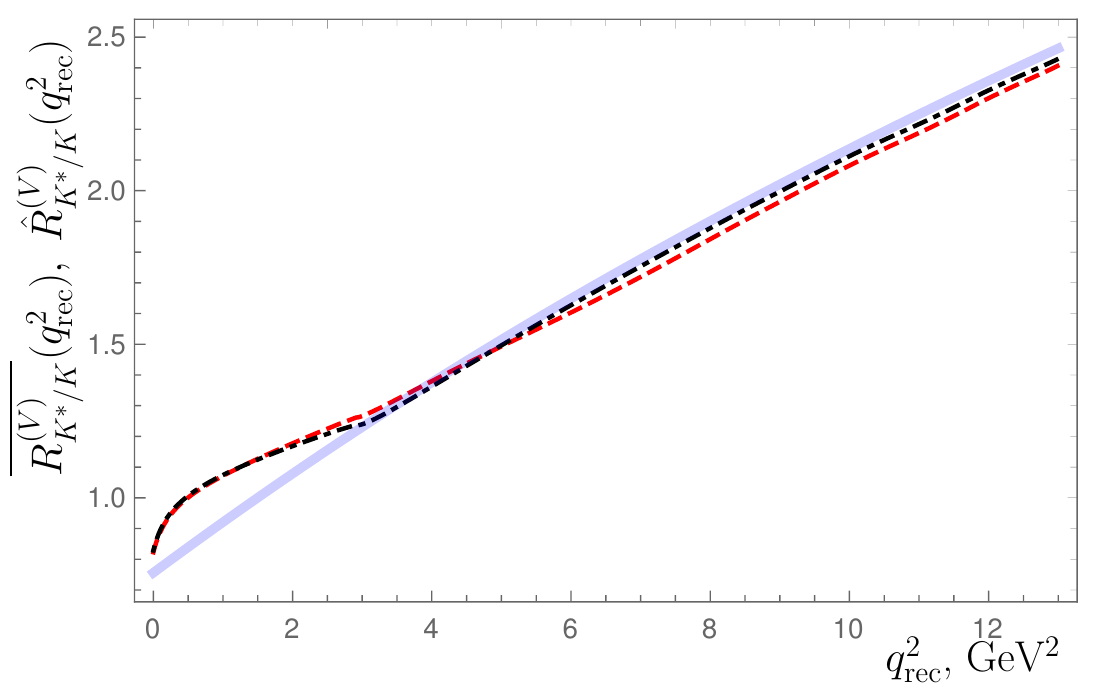}\\(a)&(b)\end{tabular}
\caption{\label{Fig:irrat}Theoretical ratio (\ref{Rdd}) [blue solid curves] vs. measurable ratio (\ref{Red}), for both~scalar (a) and~vector (b) mediators, as a function of $q_\mathrm{rec}^2$, for either our DM parameter values of Table~\ref{ParVal+} [red dashed curves]~or~a large mediator-mass sample at $M_\phi=20\mbox{ GeV}$, $\Gamma^0_\phi=20\mbox{ GeV}$ and $m_\chi=0.42\mbox{ GeV}$ for the scalar~mediator, resp.\ $M_V=20\mbox{ GeV}$, $\Gamma_V^0=20\mbox{ GeV}$ and $m_\chi=0.6\mbox{ GeV}$ for the vector mediator [black dot-dashed~curves].}\end{figure}\end{center}

The differential decay-width ratio $\widehat{\cal R}^{(R)}_{K^*/K}(q^2)$ introduced in Eq.~(\ref{Rdd}) as a function of $q^2$, which~is preferred by theory, should ``survive'' a recalculation \cite{BLM1} of the functional dependence to the~ratio $\widehat{\cal R}^{(R)}_{K^*/K}(q_\mathrm{rec}^2)$ adopting as variable $q_\mathrm{rec}^2$, which complies with all experimental conditions of Belle~II, more or less unharmed. Fortunately, these expectations are not merely wishful thinking; they are indeed realized. As established by Fig.~\ref{Fig:irrat}, the predicted ratio $\widehat{\cal R}^{(R)}_{K^*/K}(q_\mathrm{rec}^2)$ and the observed~ratio
\begin{equation}
\overline{{\cal R}^{(R)}_{K^*/K}}(q_\mathrm{rec}^2)\equiv\frac{\;\dfrac{\overline{d\Gamma^{\rm eff}}(B\xrightarrow{R}K^*\chi\bar\chi)}{dq_\mathrm{rec}^2}\;}{\dfrac{\overline{d\Gamma^{\rm eff}}(B\xrightarrow{R}K\chi\bar\chi)}{dq_\mathrm{rec}^2}}\ ,\label{Red}
\end{equation}
employing decay widths $\Gamma^{\rm eff}$ which take into consideration detection efficiencies~\cite{FGOT}, nearly~agree:
$$
\overline{{\cal R}^{(R)}_{K^*/K}}(q_\mathrm{rec}^2)\approx\widehat{\cal R}^{(R)}_{K^*/K}(q_\mathrm{rec}^2)\ .
$$
The experimental ratio (\ref{Red}) still barely depends on any DM parameter. Thus, the qualification of the differential decay-width ratios as useful tool for a discrimination of (categories of) DM~models by pinning down the spin of the bosons that mediate missing-energy excess events remains intact.

\begin{acknowledgments}
The research of A.~B.\ and D.~M.\ was carried out within the framework of the program~``Particle Physics and Cosmology of the National Center for Physics and Mathematics''.
\end{acknowledgments}

\section*{CONFLICT OF INTEREST}
All of the (three) authors of this work declare that they have absolutely no conflicts of~interest.


\begin{thebibliography}{99}
\bibitem{Belle-II:23}
I.~Adachi et al.\ (Belle II Collaboration),
{\it Evidence for $B^+\to K^+\nu\bar\nu$ decays},
Phys.~Rev.~D \textbf{109}~(2024) 112006,
arXiv:2311.14647 [hep-ex].
\bibitem{Epi24}
A.~Gaz (on behalf of the Belle and Belle II Collaborations),
{\it Recent results from Belle and Belle II},
Acta Phys.~Polon.~B Proc.~Supp.\ \textbf{17} (2024) 5-A18.
\bibitem{EPJST}
J.~Libby,
{\it Physics at Belle II},
Eur.~Phys.~J.~Spec.~Top.\ \textbf{233} (2024) 241.
\bibitem{WIFAI23}
E.~Manoni,
{\it Status and prospects for rare $B$ decays at Belle II},
PoS (WIFAI2023) 024.
\bibitem{HQL23}
R.~Volpe (on behalf of the Belle II Collaboration),
{\it Search for $B^+\to K^+\nu\bar\nu$ at Belle II},
PoS (HQL2023) 049.
\bibitem{QCD24}
Y.~Han (on behalf of the Belle and Belle II Collaborations),
{\it Recent highlights from the Belle and Belle II experiments},
Nucl.~Part.~Phys.~Proc.\ \textbf{347} (2024) 52.
\bibitem{ICHEP24}
M.~Liu (on behalf of the Belle and Belle II Collaborations),
{\it Measurements of electroweak penguin and lepton-flavour violating $B$ decays to final states with missing energy at Belle and Belle II},
PoS (ICHEP2024) 418.
\bibitem{Belle17}
J.~Grygier et al.\ (Belle Collaboration),
{\it Search for $B\to h\nu\bar\nu$ decays with semileptonic tagging~at Belle},
Phys.~Rev.~D \textbf{96} (2017) 091101(R), \textbf{97} (2018) 099902 (addendum), arXiv:1702.03224 [hep-ex].
\bibitem{BM}
A.~Berezhnoy and D.~Melikhov,
{\it $B\to K^*M_X$ vs.\ $B\to KM_X$ as a probe of a scalar-mediator dark-matter scenario},
EPL \textbf{145} (2024) 14001, arXiv:2309.17191 [hep-ph].
\bibitem{BLM1}
A.~Berezhnoy, W.~Lucha, and D.~Melikhov,
{\it Analysis of $q_{\mathrm{rec}}^2$-distribution for $B\to K M_X$ and $B\to K^* M_X$ decays in a scalar-mediator dark-matter scenario},
Phys.~Rev.~D \textbf{111} (2025) 075035, arXiv:2502.14313 [hep-ph].
\bibitem{BLM2}
A.~Berezhnoy, W.~Lucha, and D.~Melikhov,
{\it Probing vector- vs.\ scalar-mediator dark-matter scenarios in $B\to(K,K^*)M_X$ decays},
arXiv:2507.10801 [hep-ph].
\bibitem{BPR11}
B.~Batell, M.~Pospelov, and A.~Ritz,
{\it Multilepton signatures of a hidden sector in rare $B$ decays},
Phys.~Rev.~D \textbf{83} (2011) 054005, arXiv:0911.4938 [hep-ph].
\bibitem{SH13}
K.~Schmidt-Hoberg, F.~Staub, and M.~W.~Winkler,
{\it Constraints on light mediators: Confronting dark matter searches with $B$ physics},
Phys.~Lett.~B \textbf{727} (2013) 506, arXiv:1310.6752 [hep-ph].
\bibitem{GS}
G.~J.~Gounaris and J.~J.~Sakurai,
{\it Finite-width corrections to the vector-meson-dominance prediction for $\rho\to e^+e^-$},
Phys.~Rev.~Lett.~\textbf{21} (1968) 244.
\bibitem{MNNP}
D.~Melikhov, O.~Nachtmann, V.~Nikonov, and T.~Paulus,
{\it Masses and couplings of vector mesons from the pion electromagnetic, weak, and $\pi\gamma$ transition form factors}, Eur.~Phys.~J.~C~\textbf{34} (2004) 345, arXiv:hep-ph/0311213.
\bibitem{L08}
P.~Langacker,
{\it The physics of heavy $Z^\prime$ gauge bosons},
Rev.~Mod.~Phys. \textbf{81} (2009) 1199, arXiv:0801.1345 [hep-ph].
\bibitem{CMRS}
P.~Cox, A.~D.~Medina, T.~S.~Ray, and A.~Spray,
{\it Novel collider and dark matter phenomenology~of~a top-philic $Z^\prime$},
JHEP \textbf{06} (2016) 110, arXiv:1512.00471 [hep-ph].
\bibitem{HLL}
Y.~Hu, Y.~Liu, and Y.~Liu,
{\it A study on vector mediator top-philic dark matter},
Commun.~Theor.\ Phys.~\textbf{76} (2024) 085201.
\bibitem{IL}
T.~Inami and C.~S.~Lim,
{\it Effects of superheavy quarks and leptons in low-energy weak processes $K_L\to\mu\bar\mu$, $K^+\to\pi^+\nu\bar\nu$ and $K^0\leftrightarrow\bar K^0$},
Prog.~Theor.~Phys.~\textbf{65} (1981) 297, \textbf{65} (1981) 1772(E).
\bibitem{WSB}
M.~Wirbel, B.~Stech, and M.~Bauer,
{\it Exclusive semileptonic decays of heavy mesons},
Z.~Phys.~C \textbf{29} (1985) 637.
\bibitem{MS}
D.~Melikhov and B.~Stech,
{\it Weak form factors for heavy meson decays: An update},
Phys.~Rev.~D~\textbf{62} (2000) 014006, arXiv:hep-ph/0001113.
\bibitem{FL&M}
J.~A.~Bailey \textit{et al.} (Fermilab Lattice and MILC Collaborations),
{\it $B\to Kl^+l^-$ decay form factors from three-flavor lattice QCD},
Phys.~Rev.~D \textbf{93} (2016) 025026, arXiv:1509.06235 [hep-lat].
\bibitem{BSZ}
A.~Bharucha, D.~M.~Straub, and R.~Zwicky,
{\it $B\to V\ell^+\ell^-$ in the Standard Model from light-cone sum rules},
JHEP \textbf{08} (2016) 098, arXiv:1503.05534 [hep-ph].
\bibitem{KMPW}
A.~Khodjamirian, Th.~Mannel, A.~A.~Pivovarov, and Y.-M.~Wang,
{\it Charm-loop effect in $B\to K^{(*)}\ell^+\ell^-$ and $B\to K^*\gamma$},
JHEP \textbf{09} (2010) 089, arXiv:1006.4945 [hep-ph].
\bibitem{TW1}
W.~G.~Parrott, C.~Bouchard, and C.~T.~H.~Davies (HPQCD Collaboration),
{\it Standard Model predictions for $B\to K\ell^+\ell^-$, $B\to K\ell_1^-\ell_2^+$ and $B\to K\nu\bar\nu$ using form factors from $N_f=2+1+1$ lattice QCD},
Phys.~Rev.~D \textbf{107} (2023) 014511, \textbf{107} (2023) 119903(E), arXiv:2207.13371 [hep-ph].
\bibitem{TW2}
D.~Be\v cirevi\'c, G.~Piazza, and O.~Sumensari,
{\it Revisiting $B\to K^{(*)}\nu\bar\nu$ decays in the Standard Model and beyond},
Eur.~Phys.~J.~C \textbf{83} (2023) 252, arXiv:2301.06990 [hep-ph].
\bibitem{TW3}
L.~Allwicher, D.~Be\v cirevi\'c, G.~Piazza, S.~Rosauro-Alcaraz, and O.~Sumensari,
{\it Understanding the first measurement of ${\cal B}(B\to K\nu\bar\nu)$},
Phys.~Lett.~B \textbf{848} (2024) 138411, arXiv:2309.02246 [hep-ph].
\bibitem{FGOT}
K.~Fridell, M.~Ghosh, T.~Okui, and K.~Tobioka,
{\it Decoding the $B\to K\nu\nu$ excess at Belle II: Kinematics, operators, and masses},
 Phys.~Rev.~D {\bf 109} (2024) 115006, arXiv:2312.12507 [hep-ph].
\end{thebibliography}
\end{document}